\documentclass[singlecolumn,showpacs,superscriptaddress,notitlepage]{revtex4-1}
\usepackage{amsmath,amssymb,amsfonts,graphics,graphicx,dcolumn,bm}
\usepackage[toc,page]{appendix}
\usepackage{natbib}
\usepackage{geometry}
\usepackage{qtree}
\usepackage{color}

\usepackage{epstopdf}

\usepackage{multirow}

\usepackage[ddmmyyyy,hhmmss]{datetime}

\begin{document}
\title{Migration patterns across the life course of families: Gender differences and proximity with parents and siblings in Finland}

\author{Asim Ghosh}
\affiliation {Department of Computer Science, Aalto University School of Science, P.O. Box 15400, FI-00076 AALTO, Finland}
\author{Venla Berg}
\affiliation {Population Research Institute, V\"aest\"oliitto - Finnish Family Federation, Helsinki, Finland}
\author{Kunal Bhattacharya}
\author{Daniel Monsivais}
\affiliation {Department of Computer Science, Aalto University School of Science, P.O. Box 15400, FI-00076 AALTO, Finland}
\author{Janos Kertesz}
\affiliation {Center for Network Science, Central European University, Budapest, Hungary }
\affiliation {Department of Computer Science, Aalto University School of Science, P.O. Box 15400, FI-00076 AALTO, Finland}
\author{Kimmo Kaski}
\affiliation {Department of Computer Science, Aalto University School of Science, P.O. Box 15400, FI-00076 AALTO, Finland}
\author{Anna Rotkirch}
\affiliation{Population Research Institute, V\"aest\"oliitto - Finnish Family Federation, Helsinki, Finland}
\email[Corresponding author; ]{anna.rotkirch@vaestoliitto.fi}

\begin{abstract}
Family members' life course tendencies to remain geographically close to each other or to migrate due to education or job opportunities have been studied relatively little. Here we investigate migration patterns of parents and their children between 19 administrative regions of Finland from 1970 to 2012. Using the FinnFamily register dataset of 60 000 index individuals and their family members, we investigate the patterns of regional migration and regional co-residence of parents and their children. Specifically, we analyse how likely it is for children to reside in the same region as their parents at any specific age, whether parents and children who live in different regions are likely to reunite, and whether siblings function as regional attractors to each other. Results show an intense regional migration of people to the capital area. The migration propensity of individuals is high in early childhood and peaks in early adulthood. About two thirds of Finnish children live in the same region as their parents throughout their adult lives. Females show higher propensity to migrate than males, since daughters move away from their parents earlier and with a higher rate than sons do. The propensity for two full sibling brothers to be in the same region is higher than that for other types of sibling dyads. We conclude that family members serve as important geographical attractors to each other through the life course and that family attraction is stronger for sons and brothers than for daughters and sisters in contemporary Finland.
\end{abstract}

\maketitle


\section{Introduction}

In all societies, there is a well-known tendency for family members to remain close to each other throughout their lives \cite{greenwood1997migration,lundholm2015migration,madsen2007kinship,neyer2003blood,salmon2015evolutionary,stewart2008human}. The importance of family members persists also in contemporary wealthy and globalised societies. European adult parents and children provide each other substantial amounts of emotional, financial, and practical support, often on at least a weekly basis \cite{szydlik2016children} and the number of close family members has been found to be quite stable across the adult life course  \cite{wrzus2013closeness}. The persistent nature of family ties includes geographical proximity, so that adult children stay in the area of their childhood or return to it at later stages of life but also the parents may move close to their adult children in order to be with their grandchildren or receive help as they themselves age. In order to investigate family migration processes several studies, using data on residential proximity, have demonstrated that close kin live close to each other also as adults \cite{Mulder2009family}. A recent study found that in contemporary Sweden the geographical distances between adult children and their parents and grandparents remained remarkably stable once the children had reached their late 20s \cite{kolk2016life}. Most contemporary Swedes have at least some close kin within both 20 km and 75 km radius throughout their lives, and every second Swedish woman born in the 1970's is living in the same municipality as at least two of her parents or grandparents. However, comparable data on family migration across the life course is unfortunately scarce. Inspired by these findings in Sweden and by other recent studies \cite{bordone2009proximity, pers2013proximity, broek2017sibling}, with the access to appropriate family data of multiple kin generations our aim here is to investigate the migration patterns of contemporary Finns within Finland.  

This study explores kin relations as the basis for residential choices and examines the migration patterns of kin networks in order to see whether the close family members continue to act as an attractor even in the context of modernisation and urbanisation. We provide detailed analyses of the migration patterns of Finnish family members moving from one region ('maakunta' in Finnish) to another region over the years from 1970 up to 2012. Finland is known to have  experienced large migration flows from the countryside to the cities, yet we know of no other study in literature detailing this process based on the individual-level data of family members across their life course of over 40 years. Using the FinnFamily register dataset of about 60,000 index individuals and their family members, we investigate how likely it is for children to live in the same regions as their parents from birth to old age, and how likely it is for siblings to live in the same region after they have moved away from their parental homes. Our data allows us to analyse geographical proximity among both full and half siblings, which to our knowledge has not been previously done with nationally representative data. Thus our three research questions will focus on: 
(i) the regional attractiveness and migration patterns, the sending and receiving regions and net regional migration flows as well as the propensity to migrate as a function of age; 
(ii) the probability for offspring to reside in the same region as parents, and the likelihood of children and parents to "reunite" in the same region if they have at some stage been regionally separated;
(iii) the propensity of siblings to reside in the same region with each other.

\section{Data description and country background}
In this study we use the FinnFamily data, which is a representative multi-generational dataset of the late 20th century population of Finland (with the recent census of about 5.5 million), derived from the Population Register Centre of Finland. It consists of about 60,000 randomly selected Finns - considered as index-persons - from six birth cohorts of years 1955, 1960, 1965, 1970, 1975, and 1980, each having about 10,000 people constituting 11 \%  - 16 \% of the total birth cohort.  The FinnFamily data, consisting altogether 677,409 individuals, including the index persons' parents and parents' other children, i.e., siblings and half-siblings as well as the index persons' and their (half-)siblings' children and children's children. In the case of half-siblings, the data includes the half-sibling's other parent, either mother or father (randomly selected), to avoid including two half-siblings that are not genetically related. Hence the data comprise families over four generations: zeroth generation mothers and fathers of the index persons; first generation index-persons and their siblings and half-siblings; second generation cousins; and third generation second cousins. These four family generations are covered by a sequence of six birth cohorts separated by five years, which allows us to discern life course similarities across several decades. An example of a typical family structure is shown in Fig. \ref{appendix-fig1} of the Appendix. 

The data features detailed demographical information of every subject, namely the date and region of birth, the time of death, yearly information of the place of residence (i.e. region), and the time of the first five moves abroad and back to Finland. In addition, the data has information about the subjects' socioeconomic status (every five years from 1970 onwards), highest educational degree, yearly income (from 1987 onwards), and information of occupation (annually from 1987 onwards), and time of marriage and divorce, which are not used in this study. The FinnFamily data was created for the Population Research Institute at V\"aest\"oliitto by Statistics Finland.

For this study we have organised the data for the specific sub-analyses as follows. In case of the descriptive regional and migrational analysis, we consider the data of all the index persons and their family members, thus the sample size is N=674,285. To illustrate migrational flows we have adopted the flow circle method introduced in \cite{abel2014quantifying,gjabel}. For the analysis related to parents and offspring, the latter are the index persons. For the analysis of reunion, we detected in the dataset 5228 events for father-child pairs and 5851 for mother-child pairs. For the analysis concerning sibling attraction, we included only those index persons who have at least one sibling, two siblings, and three siblings with the sample size being 20556, 14306 and, 6964, respectively. There were 24113 female index persons and 25045 male index index persons with at least one full-sibling and 2851 female index persons and  2986 male index persons with at least one half-siblings.

Finland is divided into 19 administrative regions between which we will investigate  the internal migration flows. After the Second World War, during the latter half of the 20th century, the country underwent major changes and rapidly transformed from a poor and agrarian country to a wealthy and post-industrialised society. This period is characterised by urbanisation and diminishing rural population, with migration flows from the sparsely populated Northern and Eastern parts of the country to the urban centres in the Southern and Western parts. (International migration, which is not studied here, was characterised by a large emigration wave in the 1950s to more prosperous countries, especially to neighbouring Sweden, and later by a modest degree of surplus immigration as international migrants and refugees have moved to Finland).  

During our study period, educational and labour market opportunities became ever more concentrated in larger cities with simultaneous large scale educational expansion \cite{statfinland2017edu}. Specifically, women's educational level has increased strongly since the 1950s and Finnish women are currently on average more highly educated than the men are. Women's labour market participation is also strong: Finnish women typically work full time and the employment rates are quite similar between the genders, with 58\% of women and 61\% of men being employed of all the 15--74 year old Finns \cite{statfinland2017tasaarvo}. Finland has introduced several female- and family-friendly welfare state policies aimed at promoting gender equality and maternal work force participation. They include free or inexpensive health care and care for the elderly and a subjective right to public daycare for children under three years old \cite{anttonen1999childcare}. Such developments in educational levels and welfare state policies have allegedly had a major impact on the internal migration flows. On one hand, the educational expansion and urbanisation drive young people to move to university towns and families to move into larger cities or their vicinities. On the other hand, public services providing help with childcare and care for the elderly alleviate the need for kin help as only few Finns provide full-time grandchild care or full-time care of their elderly parents \cite{danielsbacka2013suku}. However, less intense form of kin help such as occasional grandparental involvement are frequent and is also highly valued by the population and  known to enhance everyday logistics and the well-being of family members \cite{nelson2014kin,coall2010grandparent}. During our study period, the median age at which Finns entered their first union (either through marriage or cohabitation) remained relatively stable at around 25 years of age \cite{jalovaara2012firstunion}. The age of having the first child rose from around approximately 25 to 28 years for women; men are typically 2 years older than women are when they become parents. The median number of children in Finland was two (2) during the whole study period and fertility rates fluctuated between 1.5 and 1.87. 


\section{Results}
In this section we present the analysis results for internal population migration between 19 regions in Finland over the four decades from 1970  to 2012. Fig. \ref{fig0} (on the left) illustrates the geographical locations of these 19 regions and the corresponding indices used to denote them. The names of the regions,  corresponding indices and the numbers of FinnFamily data index persons sampled per region are listed in Table A1 of the Appendix.


\subsection{Region-to-region migration patterns}

During the last 50 to 60 years, most regions of Finland have witnessed a sizable population movement in the form of migration towards the three main urban centres: the capital Helsinki (in Uusimaa region (m1)), the former capital Turku (in Varsinais-Suomi region (m3)) and the city of Tampere (in Pirkanmaa region (m6)). Figure 1 (on the right) and especially Figure 2 illustrate that these three regions experienced more immigration than emigration, i.e. gained in population during the study period, while other regions like Satakunta (m4), Pohjois-Pohjanmaa (m17) and Kainuu (m18) predominantly lost inhabitants.

\begin{figure}[h]
\centering
\includegraphics[width=6.5cm]{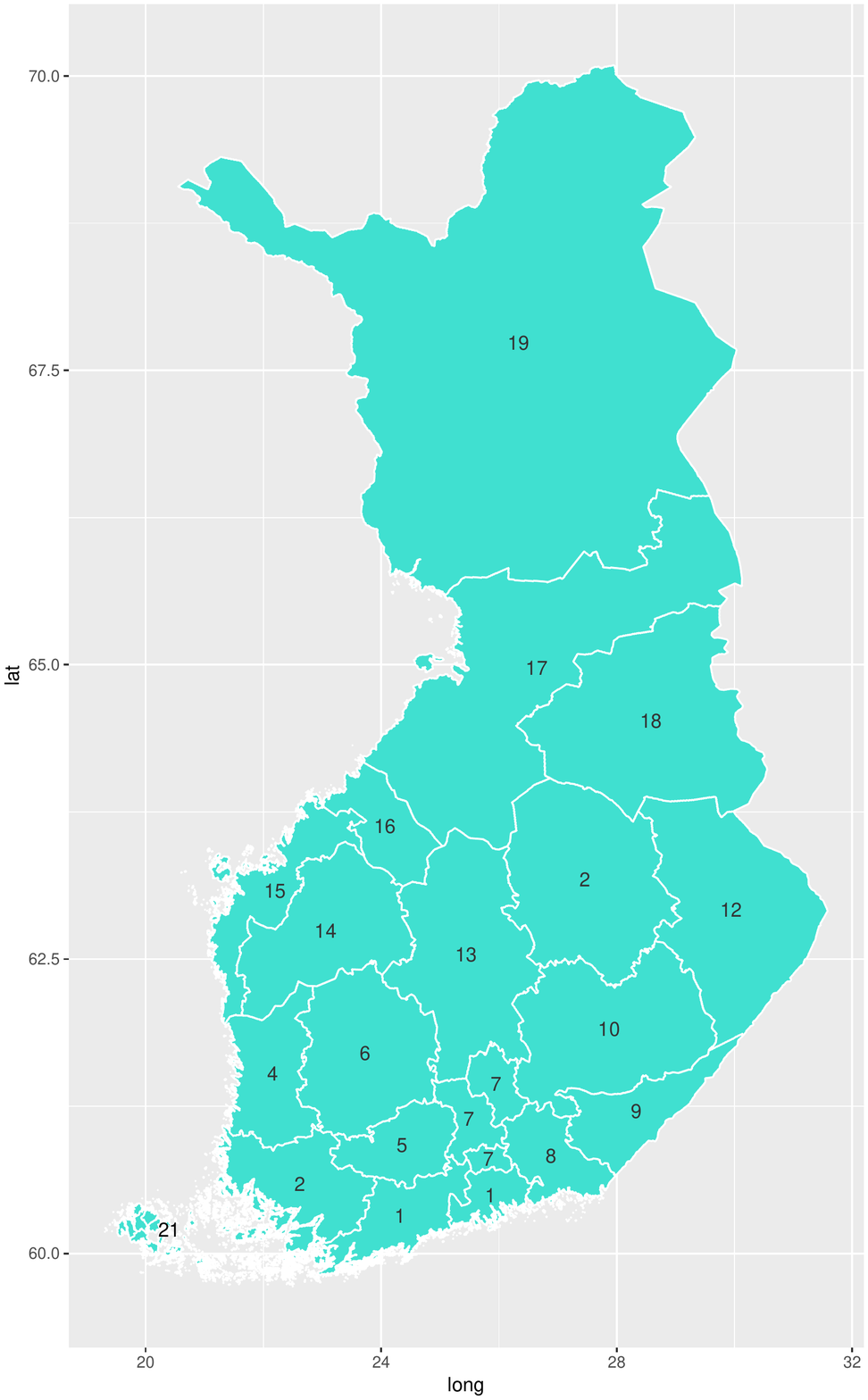}
\includegraphics[width=8.cm]{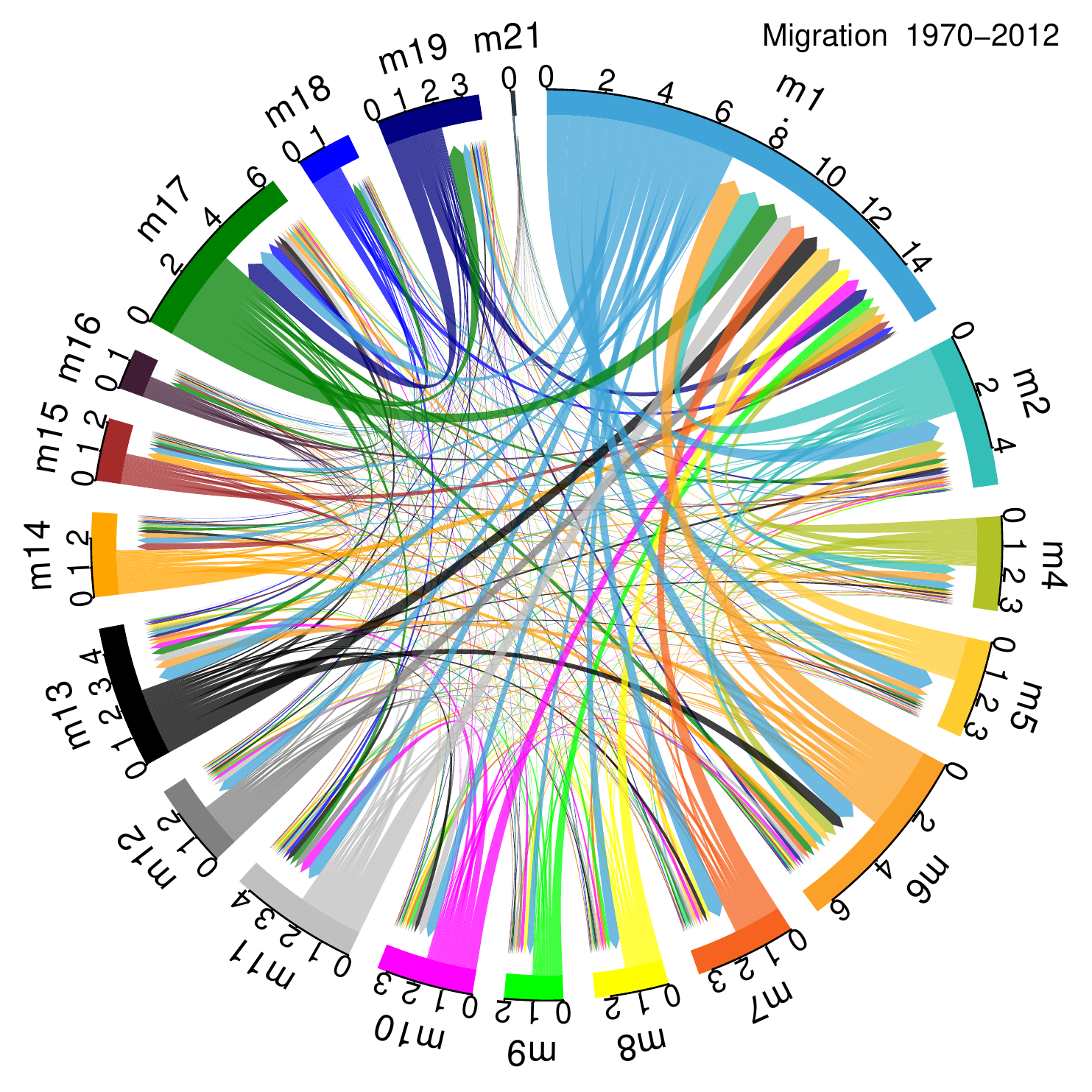}
\caption{(Left) Locations of Finland's 19 regions and corresponding labels.  (Right) Regional migration: Approximately $28\%$ of the total number of individuals in our sample participate in this migrational flow pattern. Tick marks on the circle indicate migration flows in units of 10,000 individuals.}
\label{fig0}
\end{figure}

Fig. \ref{fig0}(right) visualises the total number of individuals that moved from one region to another region over the study period, with cumulative numbers and the tick marks on the circle indicating the number of individuals in units of 10,000 (the circular plot was drawn based on the method of ref. \cite{abel2014quantifying,gjabel}). Approximately $28\%$ of the total number of individuals in the sample participate in this migrational flow. One can clearly see how crucial role the capital Helsinki in region m1 (Uusimaa) has played in the migrational dynamics of the whole country, serving as the main attractor of immigration from the other regions. When the individuals are not moving to Uusimaa region (m1), they tend to move from their region of residence to a neighbouring region. These observations are corroborated with Appendix Fig. \ref{appendix-fig3} where we show a matrix of migrational outflows from one region to another region in relative terms (\%).

In order to see whether the migration to the cities was especially prominent at some specific time period, we investigated the changes in the migration flows at two different time windows, namely 1970 - 1980 and 2000 - 2010. However, we did not observe any significant differences between these two time periods, except for the minor changes in the ranking order of inflows in the case of two regions: (i) the two largest inflows to region m4 (Satakunta) were in the ranking order from regions m6 (Pirkanmaa) followed by m2 (Varsinais-Suomi) in 1970-1980, while for 2000 - 2010 the order was reversed, i.e. m2 followed by m6 and (ii) in case of region m17 (Pohjois-Pohjanmaa) in the ranking order of m1 (Uusimaa) followed by m19 (Lappi) for 1970 - 1980 while for 2000 - 2010 the order was reversed, i.e. m19 followed by m1.
 
\begin{figure}
\begin{minipage}[b]{0.30\linewidth}
\centering
{\small
\begin{tabular}{|c|c|}
\hline
Label & Region \\ 
\hline
m1 &  Uusimaa \\
\hline
m2 & Varsinais-Suomi\\
\hline
m4 &  Satakunta \\ 
\hline
m5 & Kanta-H\"ame  \\
\hline
m6 & Pirkanmaa \\
\hline
m7 & P\"aij\"at-H\"ame  \\
\hline
m8 & Kymenlaakso \\
\hline
m9 & Etel\"a-Karjala \\
\hline
m10 & Etel\"a-Savo \\
\hline
m11 & Pohjois-Savo \\
\hline
m12 & Pohjois-Karjala \\
\hline
m13 & Keski-Suomi \\
\hline
m14 & Etel\"a-Pohjanmaa \\
\hline
m15 & Pohjanmaa \\
\hline
m16 & Keski-Pohjanmaa  \\
\hline
m17 & Pohjois-Pohjanmaa  \\
\hline
m18 & Kainuu \\
\hline
m19 & Lappi \\
\hline
m21 &  Ahvenanmaa \\
\hline
\end{tabular}
}
\par\vspace{0pt}
\end{minipage}
\begin{minipage}[b]{0.60\linewidth}
\centering
\includegraphics[width=\linewidth]{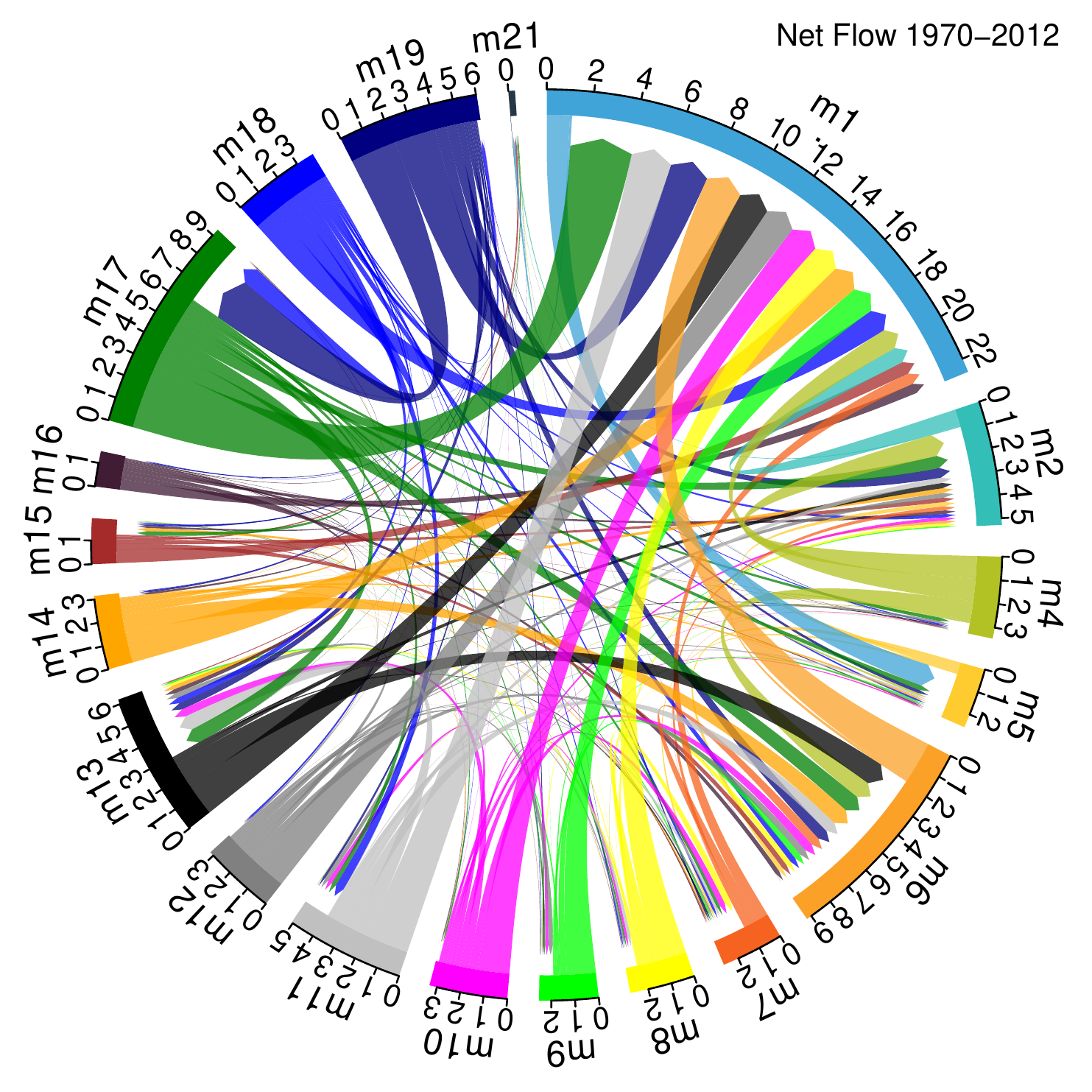}
\par\vspace{0pt}
\end{minipage}
\caption{The netflows (= outflows - inflows) of Finns between different regions. The Uusimaa (m1) region is the most prominent attractor followed by Pirkanmaa (m6) and Varsinais-Suomi (m2). Also note that Kanta-H\"ame (m5) serves as an attractor for Uusimaa (m1). Tick marks on the circle indicate net flows of people in units of 1000 individuals.}
\label{flow}
\end{figure}

Fig.\ref{flow} depicts the netflow  (= outflow - inflow) of individuals between all the pairs of regions (tick marks on the circle plot indicate the number of individuals in units of 1,000). Here we see how the Uusimaa region (m1) with the capital Helsinki is the largest attractor of migrating people followed by Pirkanmaa region (m6) with the city of Tampere and Varsinais-Suomi region (m2) with the city of Turku. Correspondingly, several regions (m4, m10, m12, m14, m16, m18, and m19) are predominantly losing their population through internal emigration, while the remaining regions (m5, m7, m8, m9, m11, m13, and m17) turn out to be quite balanced in their netflow of internal migration. Interestingly, there is a noticeable flow of people from region m1 to m5, unlike the flows between all the other regions and region m1. Hence Kanta-H\"ame region (m5) serves as an attractor for Uusimaa region (m1). This may be due to m5 being within a 100 km radius from the capital Helsinki in m1, the availability of good daily public transportation (railways and bus) services between these two regions, and cheaper house prices in Kanta-H\"ame region (m5). 
 
\subsection{Migration patterns by age}
\begin{figure}[h]
\centering
\includegraphics[width=8cm]{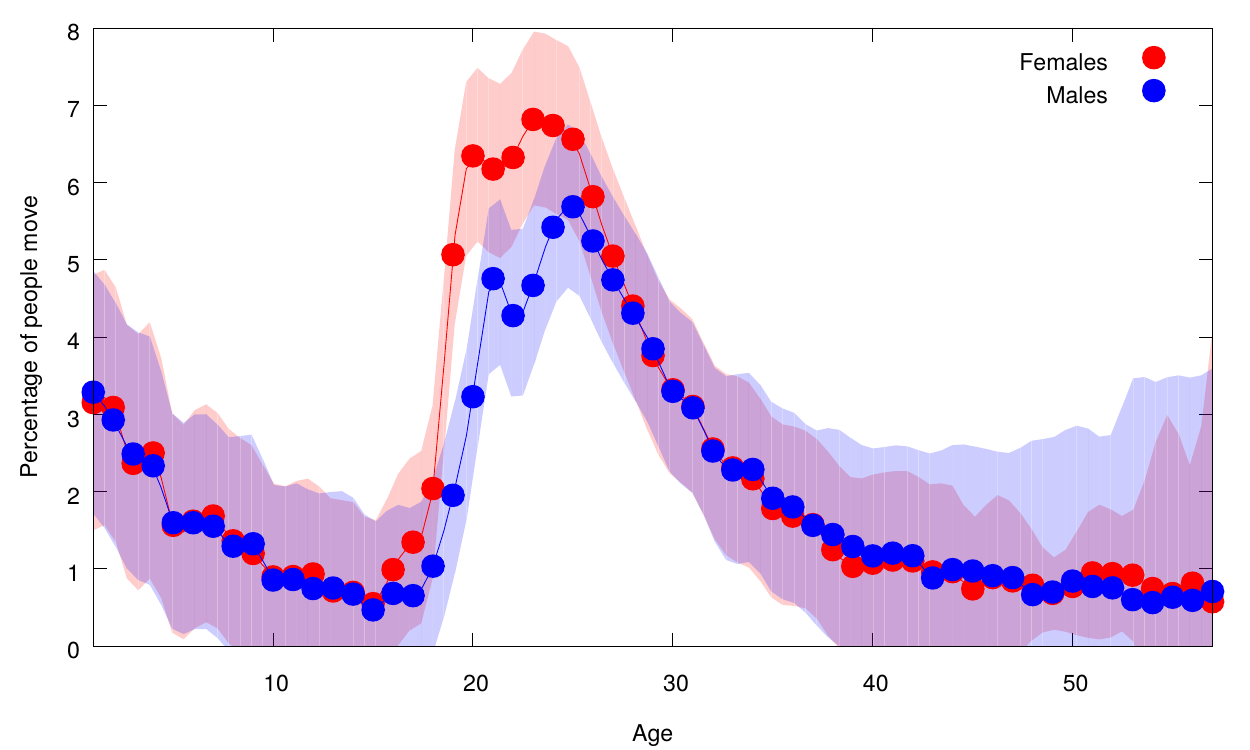}
\caption{The percentage of index individuals (sampled from six different birth cohorts) migrating from one region to another region by every year; shaded regions indicate  95\% confidence intervals. The likelihood to move peaks at infancy and at 18-28 years of age.}
\label{fig1}
\end{figure}

We now turn to investigate how the migrational patterns varies with the age of individuals. Fig. \ref{fig1} presents the likelihood of the index person to move from one region to another as a function of his or her age. We can see a strong age dependency in the migration pattern of individuals during their life course and two migrational peaks occurring at two specific stages of life. The first peak appears during the infancy of the index person, or from the birth to the first few years of life, when 2-3.3\% of index individuals were moving from one region to another, most likely together with their parents. The likelihood to migrate then declines such that Finns around the age of 16 show a very low propensity to move away and only about 0.5\% of them do so on a yearly basis. After that, the likelihood of switching region accelerates rapidly and continues to do so, peaking at the age around 26 years. This latter peak illustrates how Finns as young adults move away from the region where their parents live (and allegedly also away from their parental home) due to employment opportunities, occupational training or university studies. As we showed above, the migrants are typically moving either to the capital Helsinki in Uusimaa region (m1) or to a neighbouring region. 

As mentioned in the Data section, during the study period Finns typically became parents in their late 20s and early 30s. Hence the Figure \ref{fig1} captures the generational flow (although the studied individuals here are not related to each other): when young parents move, their young children also move. Later, when individuals are past their 30s, their likelihood to switch regions decreases, so that the yearly rate of internal migration goes down to  1\% level. Moving during the young adulthood happens a couple of years later for young men than for young women and the migration rate also appears to be higher for women. The gender differences turned out to be statistically significant from the late teen years until the late 20s, after which these differences disappear. Thus Finnish women start moving away at the age of 17 and their migration likelihood peaks at the age of about 23, when 7\% of them migrate yearly, while corresponding male migration starts at the age of 18 and accelerates less sharply to the 6\% rate and showing a peak at about the age of 25 years. 

Next, we investigate the migration patterns of parents and their children between regions. Hence our focus shifts from the migration of individuals to the regional co-residence of family member or child-parent pairs. Our second research question asks how often do the parents and children live in the same region. To investigate this, we measure the proportions of child-parent pairs, i.e. daughter-mother, son-mother, daughter-father and son-father, staying in the same region as a function of the index person's (the child's) age, see Figure 4. Results show that the majority of the Finns (i.e. two thirds) live in the same region as their parents do. Until their teens, almost all Finnish children live in the same region as their mother, and more than 95 \% live in the same region as their father does.  After that, the proportions of those who live in the same region start declining rapidly from around the age of 17 years old and subsequently stabilising around the age of 32 years old to relatively high values of more than 65\%.

In our results the gender differences in the migration patterns of young adults are also visible. For all ages, the proportions of son-mother and son-father pairs are clearly higher than the corresponding percentages for the daughter-mother and daughter-father pairs. Compared to daughters, sons at any given age are more often living in the same region as their parents. Furthermore, both women and men are more often living in the same region as their mothers than their fathers. This is not surprising, since in Finland single parents are more likely to be women, and the male lifespan is shorter than that of women. Consequently, son-mother pairs have the highest propensity to live in the same region. Of all the studied child-parent pairs, fathers and daughters are the least often sharing region of residence. However, the proportions of daughters living close to their parents increases slightly when they are middle age, while a corresponding increase is not observed for the  sons. This may indicate a `grandmother effect' of maternal grandmothers living close to their daughters as in Finland grandparents are known to provide grandchild care to their daughters and the relationship between the adult daughters and their mothers is emotionally very close \cite{danielsbacka2013suku,danielsbacka2011grandparental}.

\begin{figure}[h]
\centering
\includegraphics[width=8.5cm]{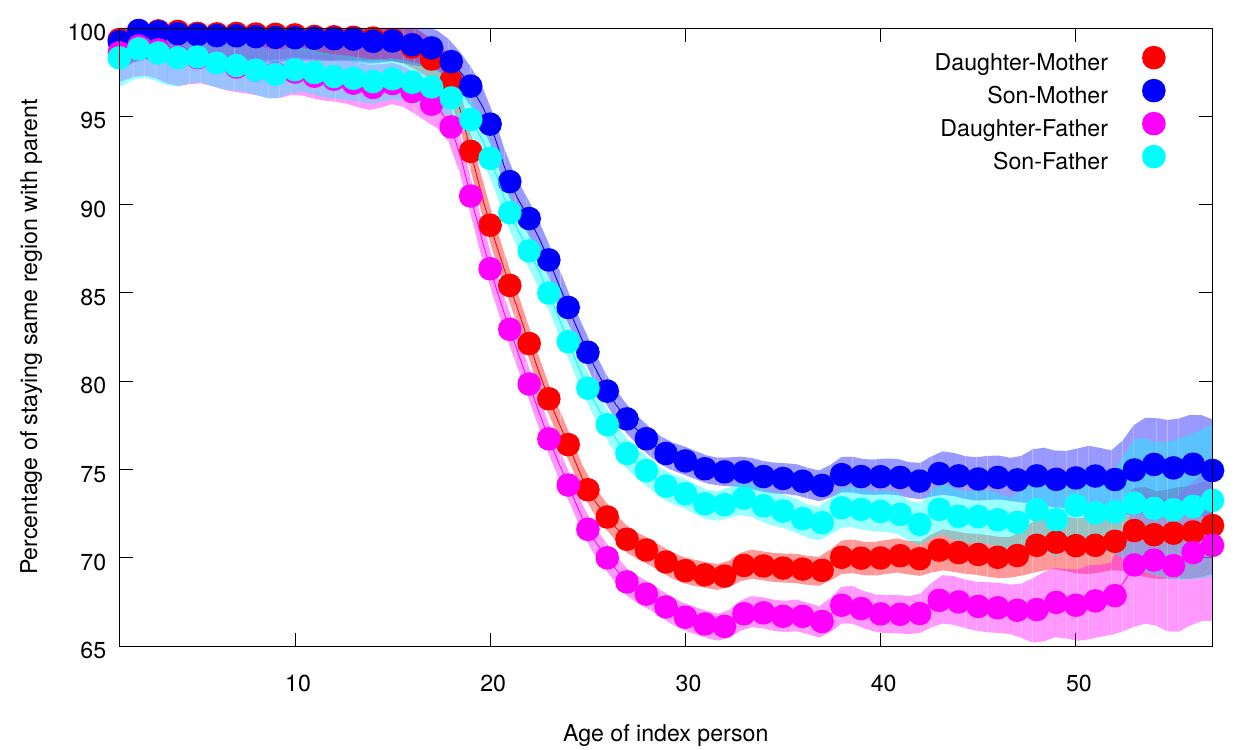}
\caption{Child-parent pairs living in the same region as a function of the index person's (i.e. the child's) age. } 
\label{fig2}
\end{figure}


\subsection{Child-parent pairs: reunion after separation}
Next, we will investigate our second research question, namely how likely are parent-child pairs reuniting in the same region following a separation. Here, we focus on those child-parent pairs in which either party has moved to a different region but later returns back to the same region. Table A2 of the Appendix features the numbers and percentages of parent - child reunions for four different child-parent pairs and for different age groups of the index person. In the FinnFamily data, we have found 5228 events of such reunion for child-father pairs and 5851 for child-mother pairs. 

\begin{figure}[h]
\centering
\includegraphics[width=10cm]{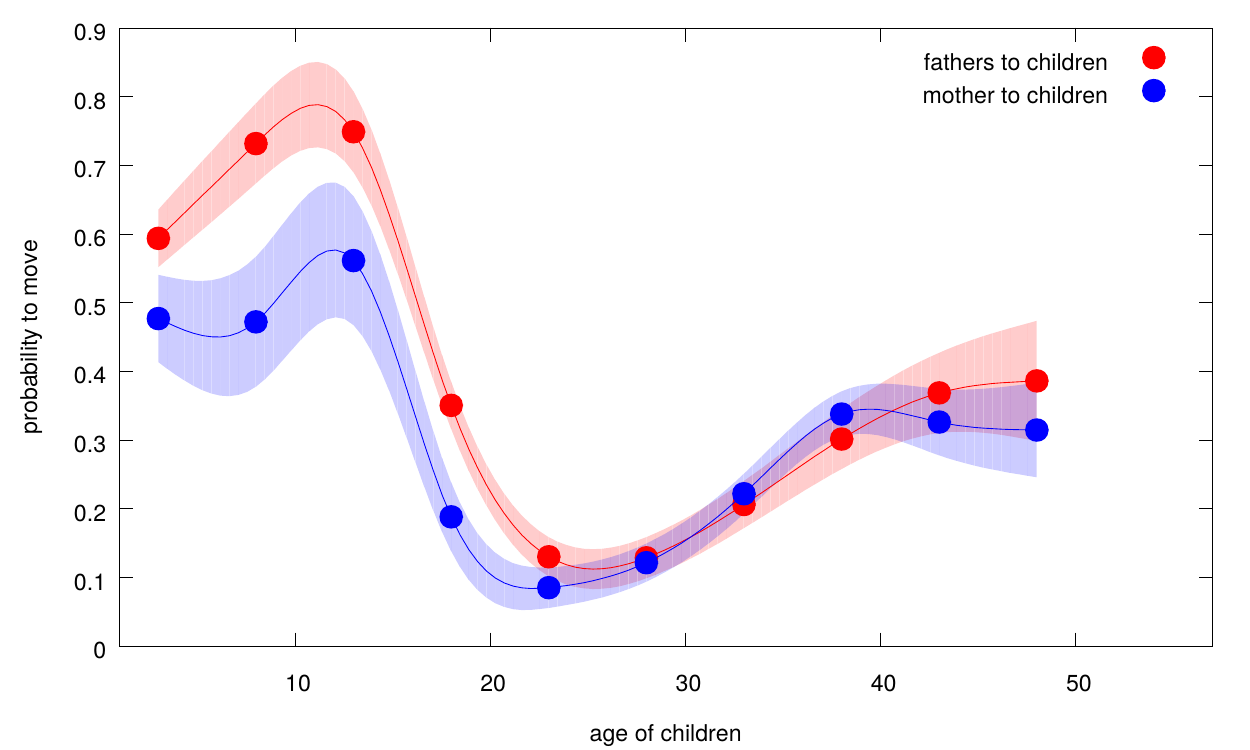}
\caption{Likelihood of reunion after separation by age of the index person for child-father and child-mother pairs, in percentages. Reunion is least likely when the child is 12-25 years old and more likely for child-father pairs compared to child-mother pairs. }
\label{fig6}
\end{figure}

Results are illustrated in Fig. \ref{fig6} which shows the relative percentages of child-parent pairs that have reunited as a function of the child's age. Until the child reaches 12 years of age, parents and children who for some reason have become residentially separated are highly likely to be reunited. The probability of reunion peaks at the age of 12 years for women and 13 years for men. During the ensuing teenage years and until the child is in his or her mid-twenties, the reunion is unlikely. It is interesting that the reunion probability declines somewhat before the general migration probability rises (at age of 17 years for women and 18 years for men, see Fig 3), especially for women. Until the early 20s, the child-father pairs are significantly more likely to reunite than the child-mother pairs are. However, fewer mothers are living separate from their children in the first place (see also Fig. 3). One can surmise that the reasons for mothers to live in another region than a young child may be quite severe, e.g. due to illness, imprisonment or child custody care, thus hindering reunion. However, more fathers than mothers would have lived in some other regions following divorce or because of work assignments, which can be considered as 'softer' reasons that would make reunion somewhat easier. Past the late teenage and early adulthood years, when the child has reached his or her late 20s and beyond, parents and children are more likely to reunite in the same region. Since most Finns had their first child in their late 20s during the study period, the result could reflect parents moving to the regions of their adult children (or vice versa) in order to facilitate provision of grandchild care. \cite{palchykov2012sex,david2016communication}. 

\subsection{Regional closeness among full and half siblings}
Our third research question concerned the possible geographical attraction the siblings represent to each other. In order to measure this effect, we calculated the proportion of an index person and one randomly chosen full or half sibling living in the same region as a function of the age of the index person. In the FinnFamily data, we found 25045 male and 24113 female index persons with at least one full sibling. In Table A3 of the Appendix we present the numbers and percentages of one sibling, two siblings, and three siblings living in the same region with the index person (their sibling) for different age groups of the index person. We have investigated four types of full sibling pairs: (i) a female index person and her sister (ii) a female index person and her brother (iii) a male index person and his sister, and (iv) a male index person and his brother. The second and third category are expected to be identical since both measure the mixed-sex sibling pairs. For all these cases, the propensity for the full siblings living in the same region is overall quite high, over 55 \% at any age. Unsurprisingly, the full siblings are very likely to live in the same region (and probably often in the same household) during their childhood. Following the main age of internal regional migration, this likelihood then decreases. We can also observe that two brothers are more often living in the same region than any other sibling pairs do. 

We have then measured the same proportions for half-siblings, i.e. for the following pairs: (i) female index person and her half-sister (ii) female index person and her half-brother (iii) male index person and his half-sister (iv) male index person and his half-brother. In the FinnFamily data we found 2986 male and 2851 female index persons with at least one half-sibling. Note that since we wanted to keep the genetical relatedness in the studied sample data it slightly underestimates the numbers of half-siblings in the population, see the Data section. 

\begin{figure}[h]
\centering
\includegraphics[width=7cm]{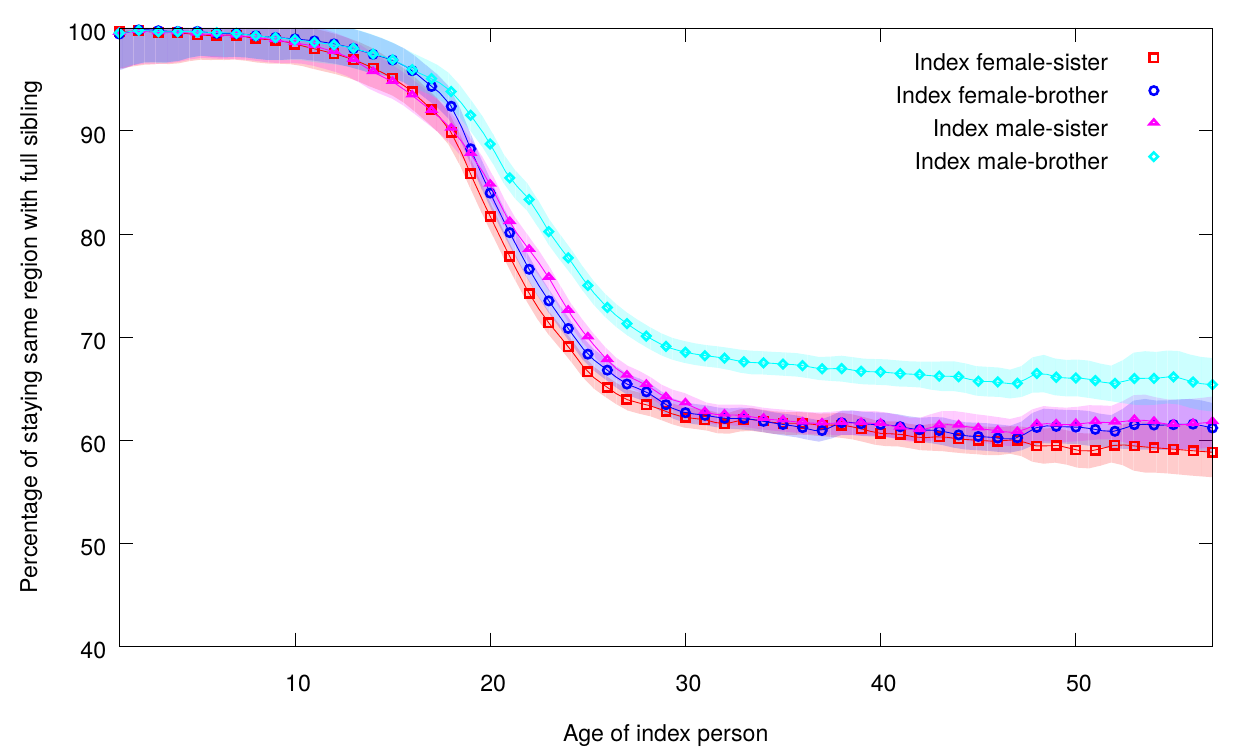}
\includegraphics[width=7cm]{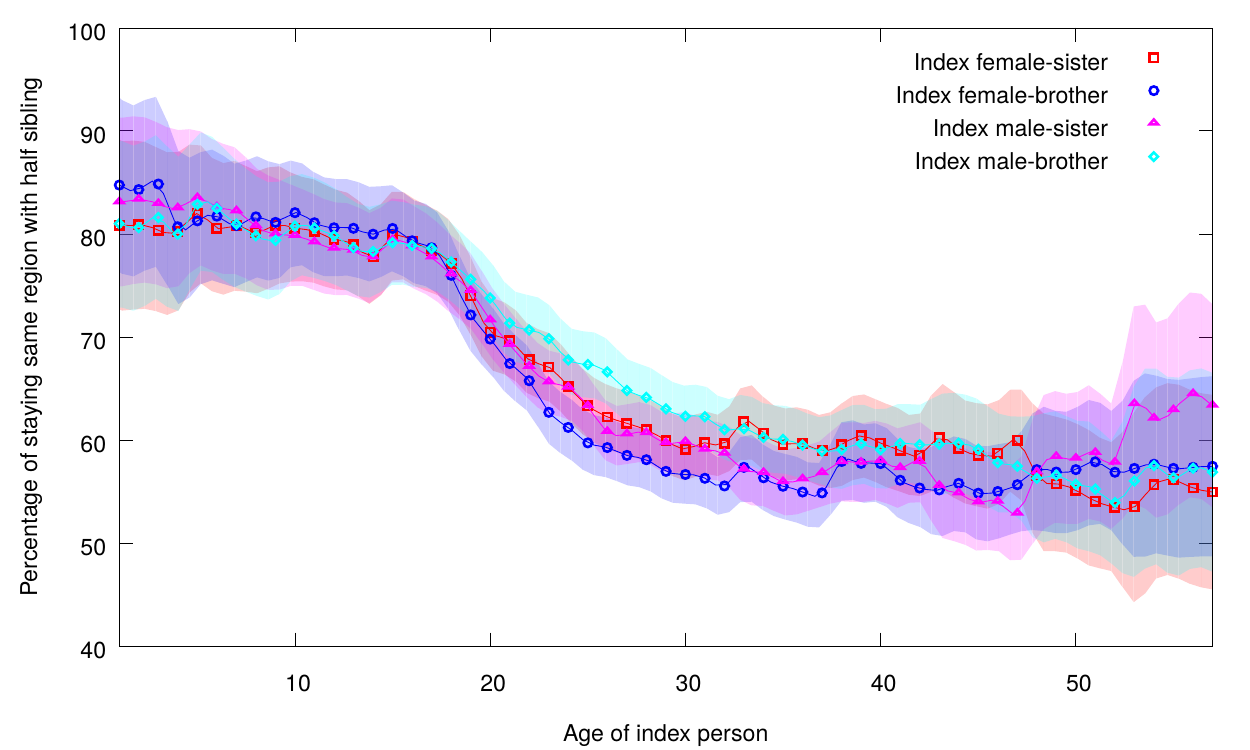}
\caption{(Left) Full siblings living in same region as another sibling by age (N females=24133, N males = 25045). Brothers most often live in the same region. (Right) Half siblings living together with another randomly chosen half-sibling by age (N females=2851, N males = 2986).}. 
\label{fig3}
\end{figure}

Fig. \ref{fig3} (right) we present the variation in the percentages of half-siblings staying in the same region as a function of the age of the index person. For all the four pairs, the propensity of an individual to live in the same region with a half-sibling decreases with age. One cannot see any statistically significant differences between different types of half-sibling pairs. In this Figure we also observe that the half siblings show overall a lower propensity to stay close to each other than full siblings do, but this difference gradually disappears after the index person has turned 30 years old. 

\begin{figure}[h]
\centering
\includegraphics[width=7cm]{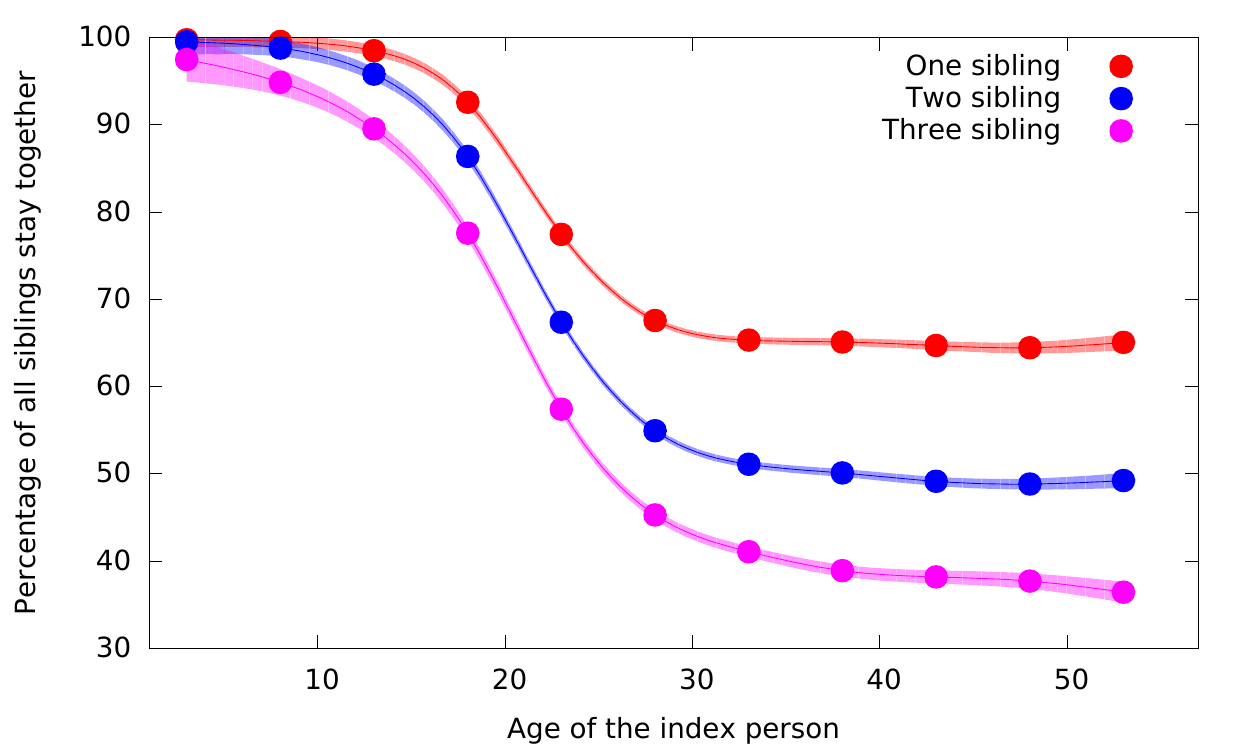}
\includegraphics[width=7.5cm]{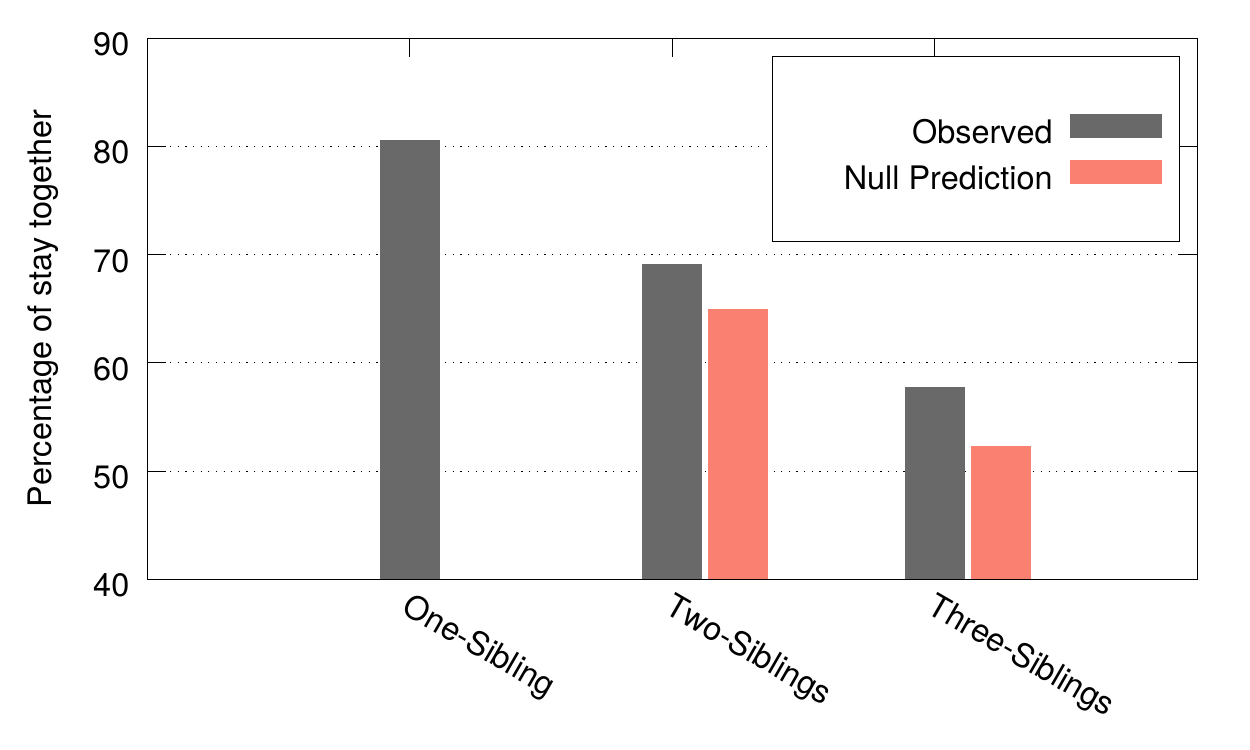}
\caption{(Left) Percentage of full siblings living in the same region as a function of the age of the index persons with one sibling (N=20556), two siblings (N=14306), and three siblings (N=6964), respectively. (Right) Percentage of full siblings living in the same region, average of yearly sum from 1970 to 2012.}
\label{fig7}
\end{figure}

Furthermore, we analyse how the sibship size affects the geographical proximity of siblings. Fig.\ref{fig7}(left) depicts the percentages of full siblings living in the same region as a function of the age of the index person having one sibling, two siblings, and three siblings. Overall, siblings are less likely to live in the same region as they age. This is obvious, since each individual has some propensity to migrate independently of the number of his or her siblings.  To know the effect of the sibship size on living in the same region, we chose the probability to stay with a single sibling as the base value $p$. The null expectation for two siblings to live in the same region as the index sibling would then be $p^2$ and for three siblings it would correspondingly be $p^3$. The average percentage for a single individual to stay in the birth region is $70\%$. Therefore, assuming no sibling attraction, the probability for finding two full siblings in the same region would be around $50\%$. This null estimation is, however, lower than the observed value, which turned out to be $65\%$ or 30\% higher than expected. Similarly, the expected null model value for two siblings and three siblings to randomly stay in the same region as the index siblings are around $35\%$ and $25\%$ but the actual results are $50\%$ and $38\%$,  respectively, or 43\% and 52\% higher than expected. Compared to the null models, the likelihood of siblings to reside in the same region is always higher than expected, and furthermore it is increasing as the numbers of full siblings increase. Based on this numerical exercise and as illustrated in Fig.\ref{fig7}(right), the observed values for sibling(s) to stay with the index person are greater than the null expectations, which indicates that siblings have a relatively high propensity to stay in the same region with each other, and that geographical sibling attraction increases with the number of siblings.

\subsection{Effect of sibling attraction on migration}
\begin{table}[h]
\caption{The assortativity coefficient, $r$ by region for sibling pairs moving to another region (every sibling has 18 other regions to migrate to). A positive value of $r$ implies that siblings attract each other, zero implies no attraction and negative values would imply repulsion.}
\begin{tabular}{|c|c|c|c|c|}
\hline
Birth  &Sibling pairs &Pairs remain together &Assortativity &Error $\sigma_r$ \\ 
place &move from &same places  &coefficient & \\ 
 & birth place & after move &r & \\ 
\hline
1 & 414 & 246 & 0.553 &  0.069 \\
\hline
2 & 235 & 136 & 0.377 & 0.094 \\
\hline
4 & 329 & 154 & 0.302 & 0.106 \\
\hline
5 & 170 & 95 & 0.298 & 0.097  \\
\hline
6 & 303 & 149 & 0.334 & 0.087 \\
\hline
7 & 222 & 119 & 0.329 & 0.089 \\
\hline
8 & 139 & 65 & 0.266 & 0.088 \\
\hline
9 & 179 & 83 & 0.242 & 0.089 \\
\hline
10 & 407 & 168 & 0.251 & 0.087 \\
\hline
11 & 470 & 214 & 0.291 &  0.084 \\
\hline
12 & 331 & 164 & 0.319 & 0.085 \\
\hline
13 & 332 & 153 & 0.311 & 0.087\\
\hline
14 & 336 & 150 & 0.314 & 0.090 \\
\hline
15 & 166 & 77 & 0.339 & 0.091 \\
\hline
16 & 113 & 42 & 0.234 & 0.088 \\
\hline
17 & 516 & 225 & 0.298 &  0.081  \\
\hline
18 & 309 & 124 & 0.275 &  0.086 \\
\hline
19 & 432 & 192 & 0.298 & 0.095 \\
\hline
\hline
\end{tabular}
\label{table1}
\end{table}

In order to quantify sibling attraction, we measure the assortativity coefficient, $r$, which denotes a preference of individuals in a social network to be connected to other similar individuals in the network. This coefficient $r$ can take values from -1 to 1, so that when $r = 1$, the network is said to be perfectly assortative, when $r = 0$ the network is devoid of such correlations, and when $r = -1$ the network is completely disassortative.

Here, we consider full siblings who were born in the same region but later migrated to another region. To calculate $r$, we counted the number of siblings pairs living in the same region after both had moved somewhere, by using the following formula for $r$ (see \cite{newman2002assortative})
\begin{eqnarray}
r=\frac{\sum \frac{e_{ii}}{N} - \sum \frac{a_{i}b_{i}}{N^2}}{1-\sum \frac{a_{i}b_{i}}{N^2}}
\end{eqnarray}
where $e_{ii}$ is the number of sibling pairs staying in the same region $i$ after siblings had migrated and $a_i$ and $b_i$ are the total numbers of observed younger individual siblings and older individual sibling for the given region $i$, respectively. The variable $i$ is summed over all the possible regions for migration and $N$ is the total number of sibling pairs moving from the given region.

Table \ref{table1} presents the assortativity coefficient, $r$, for sibling pairs for each region. We can observe that the values are positive and they turn out to be within the range from 0.23 to 0.55, which clearly indicates the existence of sibling attraction or a certain tendency for siblings to remain geographically close.  


\section{Discussion and Conclusion}

In contemporary urbanised societies family members have at different times of their lives many opportunities and incentives to migrate within their native countries. However, the patterns of these migrations have so far been studied relatively little quantitatively, mainly because of the lack of appropriate data. Here we have investigated the internal migrational patterns of parents and their children between 19 administrative regions of Finland over the period of four decades, ranging from 1970 to 2012, using the multi-generational FinnFamily register dataset of about 60,000 index individuals and their family members consisting of altogether 677,409 individuals of the country's late 20th century population. With this study we provide a first comprehensive overview of and insight into the migrational patterns of family members across their life course in terms of distinguishing variations related to age, gender and - as a unique contribution to the field - also to both full and half siblings. 

We have found that  over the 40 year period there has been strong regional migration of people to a neighbouring region or specifically to the capital Helsinki and its surroundings in Uusimaa region (m1), where now more than $25\%$ of the population of Finland is living. In our data analysis the capital and its surroundings serve as an attractor to the people of all the other regions with only one exception, the Kanta-H\"ame region (m5), to which a small net flow of people from Uusimaa region (m1) has been observed. This observation can be explained on one hand by cheaper house prices in Kanta-H\"ame region (m5) and on the other hand by good daily transportation connections to and fro the capital Helsinki in Uusimaa region (m1). 

Our analysis results for the migration patterns of individuals have indicated that the Finns' propensity to migrate is relatively high in their infancy as a result of parents moving, but even higher in their early adulthood, when young Finns between 15 and 30 years old often migrate for either their education or for work. Nevertheless, two thirds or more of Finnish children live in the same region as their parents do throughout their adult lives, like in case Sweden \cite{kolk2016life}. We have also found that daughters move away from their parents more and earlier and at a higher rate than sons do and then stay more separated from the parents. This is in line with the educational structure of the population in Finland, where a larger proportion of women than men have a degree on tertiary education \cite{statfinland2014women}. Consequently, it is more common for young women to move from their place of birth to larger cities for education, and subsequently a large proportion of those women are employed and embark on parenthood at the place of education. We believe that this is the first time the higher propensity of female than male dispersal in the contemporary Finnish population has been shown, using large and nationally representative dataset.

Apart from the educational structure in Finland, another reason for the gender difference in the migration patterns of young adults may in part be the mandatory national or military service for men, which they typically do straight after completing their secondary schooling. During their service time of about one year sons keep their domicile or home address with their parents, while daughters often leave the parental home for to enrol in higher education in another region straight after completing their secondary schooling. Consequently, a larger proportion of Finnish men will not leave their region of birth during their young adulthood. In some cases, they will remain unmarried and childless since so many young women have already left, thus creating skewed regional sex ratios for people in their 20s. In some regions of Finland the male/female- ratio is as high as 1.3-1.4 for 20-25 year old people \cite{lainiala2013sexratio}. 
 
In our analysis of the propensity of reunion after separation for different types of child-parent pairs was found to vary with the child's age and gender. We observed that when children are young, parents are likely move more to their children's regions if they had been separated for some reason than vice versa. When children themselves reach adulthood, the likelihood of parents to move to their children's and especially their daughters' places increases. This may be because the parents have tendency to move to the neighbourhoods of their children to take care of their grandchildren, which could be considered indicative of a kind of grand-parental effect in Finnish society \cite{kolk2016life,fergusson2008children}. In addition, we have found that both daughters and sons are more often living in the same region as their mothers than their fathers and that the proportions of daughters living close to their parents increases slightly when they are middle age. This in turn may indicate a `grandmother effect' by maternal grandmothers living close to their daughters, since in Finland grandparents are known to provide grandchild-care to their daughters and the relationship between the adult daughters and their mothers is emotionally very close \cite{danielsbacka2013suku,danielsbacka2011grandparental}.

Last but not least we analysed the propensity of siblings to live in the same region and found it to vary with age, gender and degree of genetic relatedness. It turned out that in general also siblings serve as regional attractors to each other across their life course. In addition we observed that two full sibling brothers were more likely to stay in the same region than the other types of sibling pairs. This is in line with the observation of males showing a low propensity to move out from their parents' regions and it may also reflect the effect of some cultural factors, like for instance a higher likelihood of brothers to become engaged in the same family enterprise than sisters. 

Based on our results we conclude that internal migrational patterns in Finland are affected by two major forces. First, the education and labour market are driving people to major cities, most importantly to the capital Helsinki and its surroundings. This pattern is evident in people generally having the highest propensity to move between regions in their early 20's and then staying more or less put, and in Uusimaa region (m1) being the biggest attractor for migrants from the rest of the country. Second, people prefer to stay close to their biological kin. Parents and their adult children have a high propensity to live in the same region, and siblings have a higher than expected probability to reside in the same region even after moving away from the region of their parents. Furthermore, the sibling attraction was observed to be higher for larger sibship sizes. In other words, siblings attract each other. 

These findings reflect Hughes's notions of the importance of kin as a basis of human residential group formation  \cite{hughes1988evolution}. Human social system relies heavily on cooperation, largely because it is exceptionally demanding to raise the offspring \cite{hrdy2011mothers}. In this biological and affinal kin are more likely to provide help, because helping biologically related individuals is beneficial in terms of ones own fitness \cite{hamilton1964genetical}. For example, help provided to biological siblings seems to be more automatic and less based on reciprocity than help provided to close non-kin friends \cite{rotkirch2014gratitude}. Thus, residing in close proximity to a biological kin seems to be of interest to modern humans, throughout their lifespan. According to our results, individuals seem to balance their educational and occupational aspirations with the need to be close to parents and siblings.

We conclude that adult family members continue to serve as important geographical or regional attractors to each other across the life course even in a wealthy contemporary society like Finland, and that family attraction is stronger for sons and brothers than for daughters and sisters.


\section*{Acknowledgements}
A.G., D.M., K.B., and K.K. acknowledges financial support by the Academy of Finland Research project (COSDYN) No. 276439 and EU HORIZON 2020 FET Open RIA project (IBSEN) No. 662725. J.K. was partially supported by H2020 FETPROACT-GSS CIMPLEX Grant No. 641191.

\newpage

\section*{Appendix}
\renewcommand{\thefigure}{A\arabic{figure}}
\setcounter{figure}{0}
\renewcommand{\thetable}{A\arabic{table}}
\setcounter{table}{0}

\begin{figure}[h]
\centering
\includegraphics[width=10.5cm]{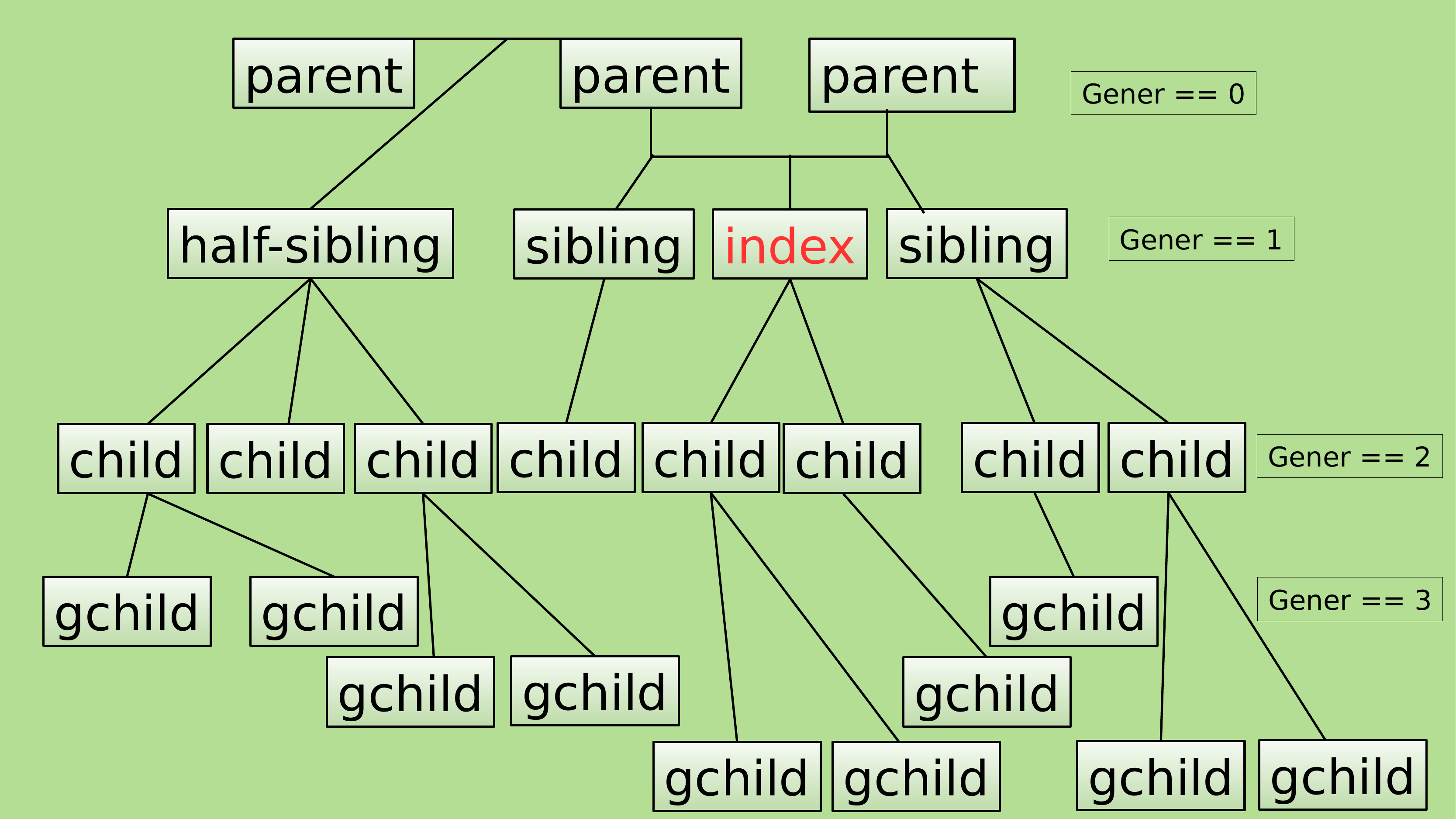}
\caption{The structure of data associated with every individual index person including individual level information from his or her parents to all genetically related offspring and offspring's offspring. Note that the data set includes information of the index person's half-siblings.}
\label{appendix-fig1}
\end{figure}

\begin{figure}[h]
\centering
\includegraphics[width=15.5cm]{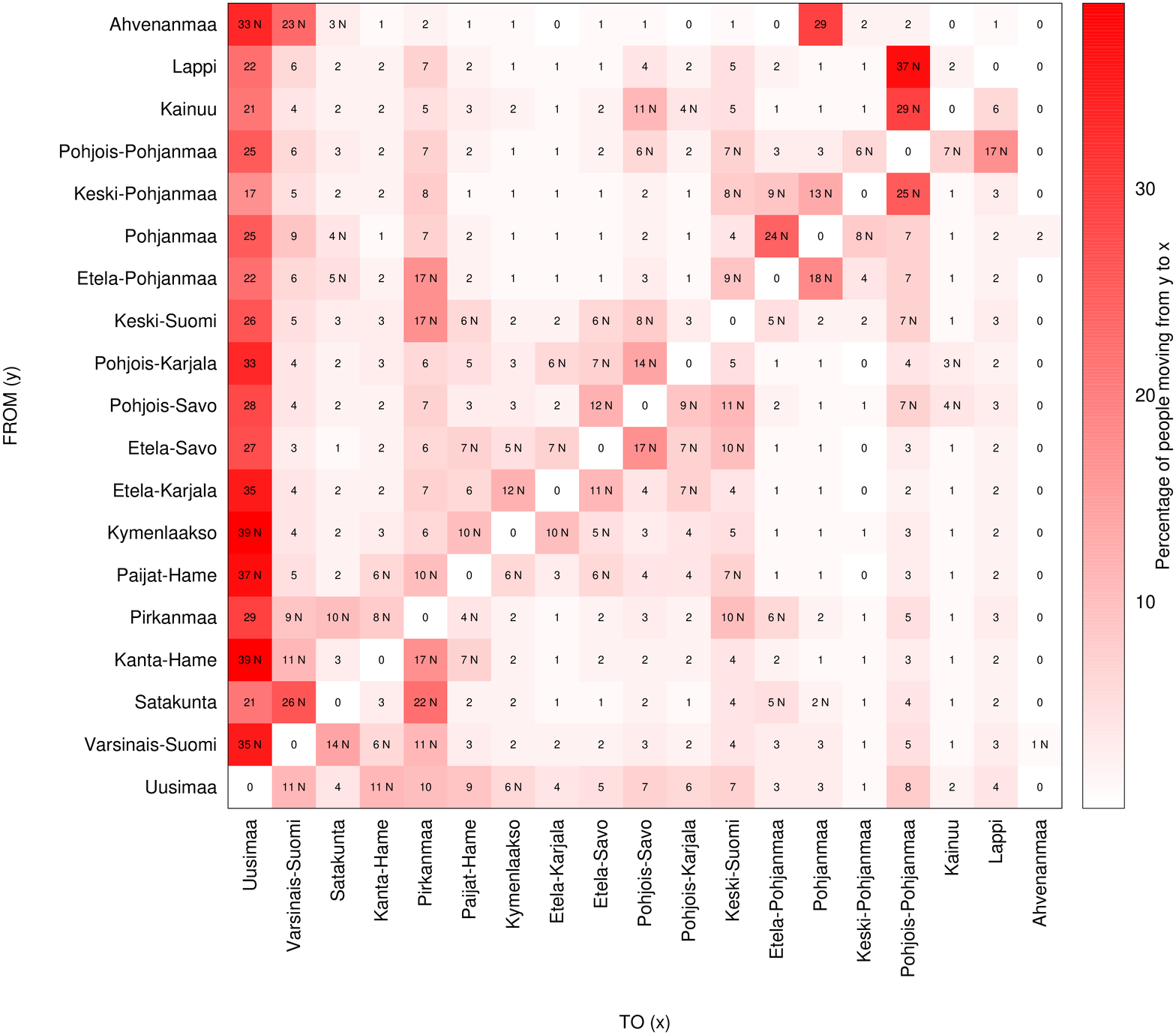}
\caption{The percentage of people moving from one region to another region. The total outflow is scale to 100\% for each region such that each row adds up to 100. It is notable that a large fraction of people from different regions go to region m1 (Uusimaa) such that few regions are getting a large fraction of people migrating from neighbouring regions (see for example region m6 (Pirkanmaa) and region m17 (Pohjois-Pohjanmaa)). With extra N in the box we indicate if two are neighbouring each other. }
\label{appendix-fig3}
\end{figure}

The FinnFamily data set consists of data of about 60,000 randomly selected Finns, named as index-persons, from six birth cohorts of years 1955, 1960, 1965, 1970, 1975, and 1980, each with about 10,000 people, and index persons' parents and parents' other children, i.e., siblings and half-siblings as well as the index persons' and their (half-)siblings' children and children's children. In the case of half-siblings, the data includes the half-sibling's other parent, either mother or farther (randomly selected), to avoid including two half-siblings that are not genetically related. Thus the data structure depicted in Fig. \ref{appendix-fig1} represents a part of the index person's family tree with his or her mother and father, siblings and half-siblings; cousins; and second cousins. In Table A1 we present the labelling and the names of the 19 regions with the number of randomly selected index persons in each region. In Fig. \ref{appendix-fig3} we show a matrix of migrational outflows in relative terms (\%) from one region to another. In Table A2  we show the numbers and percentages of reunions or 'who moves to whom' for different child - parent pairs after having lived in a different region but later moved to the same region for different age groups of the index person (child). In Table A3  we show the numbers and percentages of one sibling, two siblings, and three siblings living in the same region with the sibling index person for different age groups of the index person.

\begin{table}[h]
\caption{The labelling and names of the 19 regions of Finland with the number of randomly sampled index persons in each region.}
\label{A1}
\begin{tabular}{|c|c|c|}
\hline
Label & Region & Index persons  \\ 
\hline
m1 &  Uusimaa & 10851 \\
\hline
m2 & Varsinais-Suomi & 3320 \\
\hline
m4 &  Satakunta & 2489\\ 
\hline
m5 & Kanta-H\"ame & 1443 \\
\hline
m6 & Pirkanmaa & 3723 \\
\hline
m7 & P\"aij\"at-H\"ame & 1818 \\
\hline
m8 & Kymenlaakso & 1080 \\
\hline
m9 & Etel\"a-Karjala & 1299 \\
\hline
m10 & Etel\"a-Savo & 1872  \\
\hline
m11 & Pohjois-Savo & 2984 \\
\hline
\end{tabular}
\begin{tabular}{|c|c|c|}
\hline
level & region  &  index persons \\ 
\hline
m12 & Pohjois-Karjala & 1929 \\
\hline
m13 & Keski-Suomi & 2426\\
\hline
m14 & Etel\"a-Pohjanmaa & 2131\\
\hline
m15 & Pohjanmaa & 1580\\
\hline
m16 & Keski-Pohjanmaa & 715 \\
\hline
m17 & Pohjois-Pohjanmaa &  4226 \\
\hline
m18 & Kainuu & 1327 \\
\hline
m19 & Lappi & 2364 \\
\hline
m21 &  Ahvenanmaa & 251 \\
\hline
&  \\

\hline
\end{tabular}
\end{table}

\begin{table}[h]
\caption{Reunion of child -parent pairs: the number and percentages of reunions or `who moves to whom' after having lived in a different region and then reuniting for different age groups of sampled over all index persons.}
 \begin{tabular}{|l||l|l|l|l|}
\hline
Age group &\multicolumn{4}{l|}{parents move to children's place/total cases between the pair  (percentage with 95\% confidence)} \\
\cline{2-5}
 & Mothers to daughters  & Mothers to sons & Fathers to daughter & Fathers to sons  \\
\hline

1-5 & 49/116 (42.2 $\pm$ 9.4) & 65/123 (52.8 $\pm$ 8.9) & 167/286 (58.4 $\pm$ 6.0) & 191/317 (60.3 $\pm$ 5.8) \\
6-10 & 15/43 (34.9 $\pm$ 16.9) & 36/65 (55.4 $\pm$ 12.4) & 158/217 (72.8 $\pm$ 8.5) & 183/249 (73.5 $\pm$ 8.0) \\
11-15 & 29/54 (53.7 $\pm$ 13.4)  & 35/60 (58.3 $\pm$ 13.2) & 190/255 (74.5 $\pm$ 8.1) & 171/227 (75.3 $\pm$ 8.7)\\
16-20 & 77/512 (15.0 $\pm$ 6.8) & 79/316 (25.0 $\pm$ 7.3) & 174/609 (28.6 $\pm$ 4.9) & 185/415 (44.6 $\pm$ 4.9) \\
21-25 & 160/1912 (8.4 $\pm$ 3.9) & 127/1453 (8.7 $\pm$ 4.5) & 199/1755 (11.3 $\pm$ 3.9) & 206/1355 (15.2 $\pm$ 4.2)\\
26-30 & 213/1735 (12.3 $\pm$ 3.9)  & 198/1644 (12.0 $\pm$ 4.0)& 182/1436 (12.7 $\pm$ 4.2) & 182/1391 (13.1 $\pm$ 4.3) \\
31-35 & 264/1117 (23.6 $\pm$ 4.0) & 224/1081 (20.7 $\pm$ 4.2) & 169/843 (20.0 $\pm$ 4.9) & 173/816 (21.2 $\pm$ 4.8) \\
36-40 & 204/560 (36.4 $\pm$ 4.6) & 187/597 (31.3 $\pm$ 4.8) & 102/353 (28.9 $\pm$ 6.5) & 122/390 (31.3 $\pm$ 5.9) \\
41-45 & 104/281 (37.0 $\pm$ 6.4) & 78/277 (28.2 $\pm$ 7.4)& 59/161 (36.6 $\pm$ 8.5) & 66/178 (37.1 $\pm$ 8.0) \\
46-50 &  54/146 (37.0 $\pm$ 8.9) & 36/140 (25.7 $\pm$ 10.8)  & 28/69 (40.6 $\pm$ 12.4) & 28/76 (36.8 $\pm$ 12.4) \\
\hline
\hline
\end{tabular}
\end{table}

\begin{table}[h]
\caption{ The numbers and percentages of one sibling, two siblings, and three siblings living in the same region with the sibling index person for different age groups of the index person.}
 \begin{tabular}{|l||l|l|l|}
\hline
Age group &\multicolumn{3}{l|}{\# of siblings stay all together/out of \# (percentage)} \\
\cline{2-4}
 & One sibling  & Two siblings & Three siblings  \\
\hline

1-5 & 47096/47226 (99.72 $\pm$ 0.9) & 17235/17326 (99.47 $\pm$ 1.5) & 5528/5672 (97.46 $\pm$ 2.5) \\
6-10 &  76235/76596 (99.52 $\pm$ 0.7) & 38684/39163 (98.77 $\pm$ 1.0) & 13582/14320 (94.84$ \pm$ 1.5) \\
11-15 & 90471/ 91874 (98.4729 $\pm$ 0.6) &  55739/58195 (95.77 $\pm$ 0.8) & 22442/25070 (89.51$ \pm$ 1.0) \\
16-20 & 93376/100887 (92.55 $\pm$ 0.5) & 59143/68486 (86.35 $\pm$ 0.6)  & 25002/32231 (77.57$ \pm$ 0.8) \\
21-25 & 76698/99061 (77.42 $\pm$ 0.4) & 45278/67202 (67.37 $\pm$ 0.4) & 18196/31703 (57.39$ \pm$ 0.6) \\
26-30 & 65742/97289 (67.57 $\pm$ 0.4) & 36114/65748 (54.92 $\pm$ 0.4) & 14071/31064 (45.29 $\pm$ 0.6) \\
31-35 &  55408/ 84839 (65.30 $\pm$ 0.4) & 29615/57963 (51.09 $\pm$ 0.4) & 11590/28209 (41.08 $\pm$ 0.6) \\
36-40 & 42156/64761 (65.09 $\pm$ 0.4) & 23350/46611 (50.09 $\pm$ 0.5) & 9376/24094 (38.91  $\pm$ 0.7 ) \\
41-45 & 29167/ 45084 (64.69 $\pm$ 0.5) & 17502/35616 (49.14 $\pm$ 0.5) & 7560/19797 (38.18 $\pm$ 0.8) \\
46-50 & 17949/27853 (64.44 $\pm$ 0.7) & 11836/24241 (48.82 $\pm$ 0.6) & 5430/14399 (37.71 $\pm$ 0.9) \\
51-55 & 9406/14459 (65.05 $\pm$ 0.9) & 6429/13060 (49.22 $\pm$ 0.9) & 2985/8192 (36.43 $\pm$ 1.2) \\
\hline
\hline
\end{tabular}
\end{table}


\end{document}